\documentclass[ reprint,
 amsmath,amssymb,
 aps,twocolumn,superscriptaddress]{revtex4-1}

\usepackage[nopar]{lipsum}
\usepackage{graphicx}
\usepackage{dcolumn}
\usepackage{bm}
\usepackage[capitalize]{cleveref}
\let\revappendix\appendix

\usepackage{siunitx}

\begin{document}

\preprint{APS/123-QED}

\title{Nonlinear response of 2DEG in the quantum Hall regime}

\author{Shuichi Iwakiri}
\email{siwakiri@phys.ethz.ch}
\affiliation{
 Department of Physics, ETH Z\"urich, Otto-Stern-Weg 1, 8093 Z\"urich, Switzerland
}
\author{Lev V. Ginzburg}
\affiliation{
 Department of Physics, ETH Z\"urich, Otto-Stern-Weg 1, 8093 Z\"urich, Switzerland
}
\author{Marc P. Röösli}
\affiliation{
 Department of Physics, ETH Z\"urich, Otto-Stern-Weg 1, 8093 Z\"urich, Switzerland
}

\author{Yigal Meir}
\affiliation{
 Department of Physics, Ben-Gurion University of the Negev, Beer-Sheva, 84105 Israel
}

\author{Ady Stern}
\affiliation{
 Department of Condensed Matter Physics, Weizmann Institute of Science, Rehovot, Israel
}
\author{Christian Reichl}
\affiliation{
 Department of Physics, ETH Z\"urich, Otto-Stern-Weg 1, 8093 Z\"urich, Switzerland
}
\author{Matthias Berl}
\affiliation{
 Department of Physics, ETH Z\"urich, Otto-Stern-Weg 1, 8093 Z\"urich, Switzerland
}
\author{Werner Wegscheider}
\affiliation{
 Department of Physics, ETH Z\"urich, Otto-Stern-Weg 1, 8093 Z\"urich, Switzerland
}
\author{Thomas Ihn}
\affiliation{
 Department of Physics, ETH Z\"urich, Otto-Stern-Weg 1, 8093 Z\"urich, Switzerland
}
\author{Klaus Ensslin}
\affiliation{
 Department of Physics, ETH Z\"urich, Otto-Stern-Weg 1, 8093 Z\"urich, Switzerland
}

\date{\today}

\begin{abstract}
Breaking of inversion symmetry leads to nonlinear and nonreciprocal electron transport, in which the voltage response does not invert with the reversal of the current direction. Many systems have incorporated inversion symmetry breaking into their band or crystal structures. In this work, we demonstrate that a conventional two-dimensional electron gas (2DEG) system with a back gate shows non-reciprocal behavior (with voltage proportional to current squared) in the quantum Hall regime, which depends on the out-of-plane magnetic field and contact configuration. The inversion symmetry is broken due to the presence of the back gate and magnetic field, and our phenomenological model provides a qualitative explanation of the experimental data. Our results suggest a universal mechanism that gives rise to non-reciprocal behavior in gated samples.
\end{abstract}
\maketitle

\section{Introduction}
The current-voltage ($IV$) characteristics of a conductor can be generally expressed as $V=RI+R^{(2)}I^2+R^{(3)}I^{3}\cdots$, where $R$, $R^{(2)}$, and $R^{(3)}$ are the linear and nonlinear resistances.
The linear response ($V\propto I$) is understood in general frameworks such as the linear response theory\cite{Kubo1957} and Landauer-B\"uttiker theory\cite{Landauer1957,Landauer1970,Buttiker1986}, while the understanding of the nonlinearities is still a challenge. Nonlinear responses can be categorized as reciprocal or non-reciprocal depending on whether the voltage response $V$ switches sign upon reversal of current direction ($I\rightarrow-I$). The former typically arises due to time and/or spatial symmetries \cite{Casimir1945,Onsager1931,Onsager21931} whereas the latter occurs when those symmetries are broken.
While non-reciprocal responses have been observed in several systems such as magneto-chiral effect\cite{Rikken2001,Tokura2018}, superconducting diode effect\cite{Ando2020,Daido2022}, and nonlinear Hall effect\cite{Sodemann2015,Ma2019}, they have often been attributed to peculiar band or crystal structures. 


In this work, we demonstrate non-reciprocal behavior in the quantum Hall regime of a conventional system of two-dimensional electron gas (2DEG) with a back gate. Specifically, we measure the linear ($V\propto I$) and the lowest-order non-reciprocal ($V\propto I^2$) response of a GaAs/GaAlAs 2DEG, which was observed only for a device with a back gate. We also investigate the symmetry of the observed non-reciprocity with respect to magnetic field and contact configuration. Our observations are explained qualitatively by a model that describes the spatial modulation of carrier density due to the application of a bias current and the capacitive coupling between the back gate and the 2DEG.

\section{Experiment}
\begin{figure*}
\begin{center}
\includegraphics[width=\textwidth]{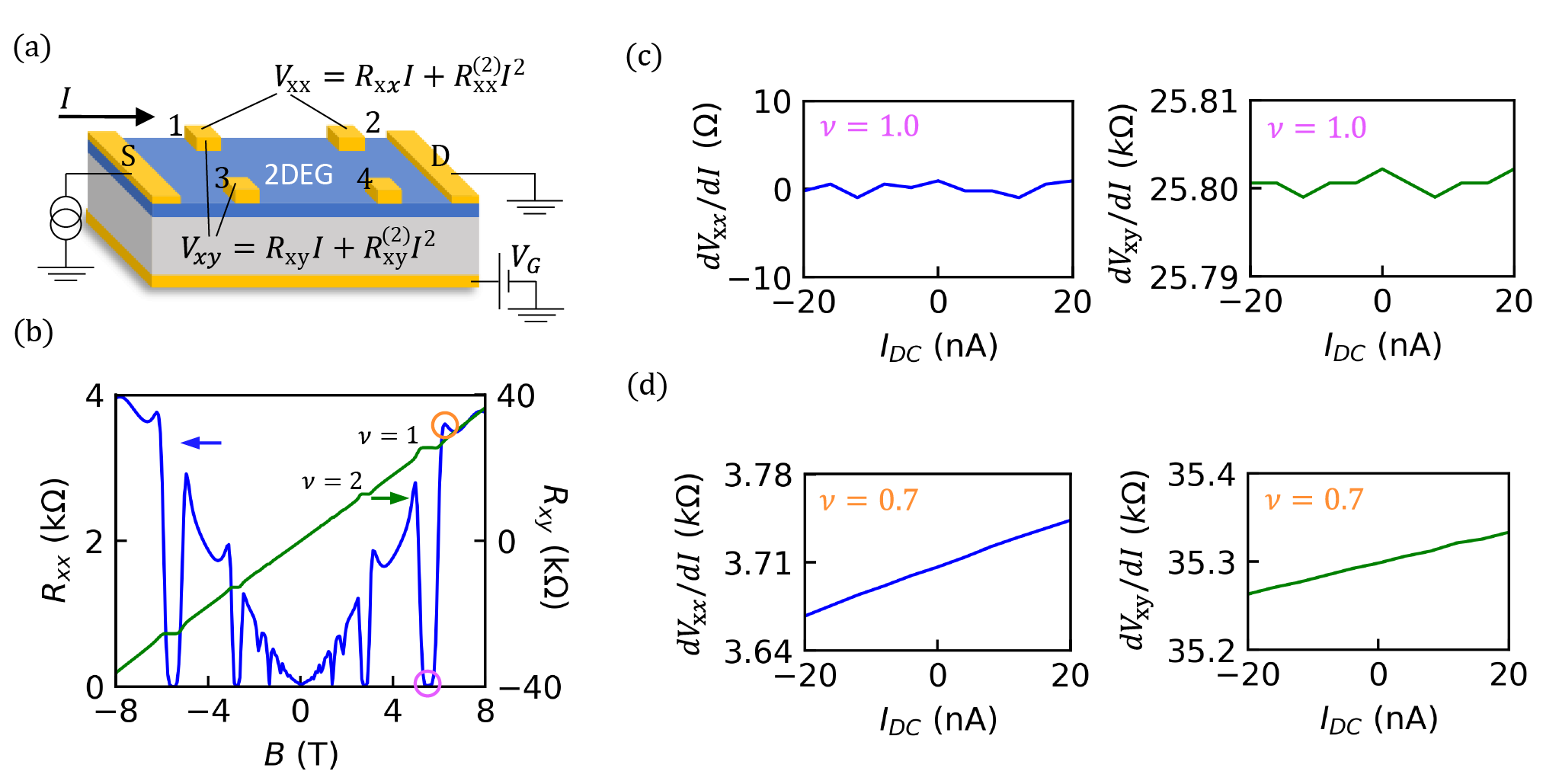}
\caption{(a) Schematic of the device. 2DEG, source and drain electrode (S and D), four voltage contacts to measure $V_{xx}$ and $V_{xy}$ (1 to 4), and back gate are shown. (b) Magnetic field dependence of the linear longitudinal ($R_{xx}$, blue) and transverse resistance ($R_{xy}$, green). (c),(d) DC current $I_{\textrm DC}$ dependence of the differential resistances in plateau ($\nu=1$) regime ((c)left panel: $\frac{dV_{xx}}{dI}$, right panel: $\frac{dV_{xy}}{dI}$) and outside plateau ($\nu=0.7$) regime ((d)left panel: $\frac{dV_{xx}}{dI}$, right panel: $\frac{dV_{xy}}{dI}$).}
\label{sample}
\end{center}
\end{figure*}

The experimental setup is shown schematically in Figure \ref{sample}(a). We used a GaAs/GaAlAs heterostructure that hosts a 2DEG buried 200 nm below the surface. A back gate with voltage $V_{\textrm G}$, located 1 $\mu$m below the 2DEG, allowed us to vary the electron density from 1.5$\times$10$^{11}$/cm$^2$ to 2.7$\times$10$^{11}$/cm$^2$\cite{Berl2016}. Au/Ge Ohmic contacts were attached to the 2DEG to inject a source-drain current and measure the longitudinal ($V_{xx}$) and transverse ($V_{xy}$) voltage response. The sample was patterned into a Hall bar shape, where the distance between the source and drain contacts, the two contacts measuring $V_{xy}$, and the two contacts measuring $V_{xx}$ are 1500 $\mu$m, 800 $\mu$m, and 400 $\mu$m, respectively. All measurements were performed at 60 mK. Carrier density was measured using the classical Hall effect at low magnetic fields ($<0.3$ T).

The current voltage characteristics of a 2DEG with the lowest-order non-reciprocal term is written as $V_{xx}=R_{xx}I+\frac{1}{2}R_{xx}^{(2)}I^2$ and $V_{xy}=R_{xy}I+\frac{1}{2}R_{xy}^{(2)}I^2$ respectively. Here, the bias current is small enough to let us ignore the component whose order is higher than $I^3$. 
The central goal of this Article is to investigate the existence and the behavior of the non-reciprocal response ($R_{xx}^{(2)}$ and $R_{xy}^{(2)}$) of gated 2DEG.
First, we measured linear response ($R_{xx}$ and $R_{xy}$) by applying an AC source-drain current $I_{\textrm{AC}}$ with a lock-in amplifier at frequency $f=27$ Hz and amplitude 5 nA. The back gate voltage is fixed at $V_{\textrm{G}}=-2$ V, making carrier density 1.35$\times$10$^{11}$/cm$^2$.
Figure \ref{sample}(b) shows the out-of-plane magnetic field ($B$) dependence of $R_{xx}$ and $R_{xy}$. As seen in the plateau of $R_{xy}$ and the zero of $R_{xx}$, the integer (filling factor $\nu=1,2,...$) quantum Hall effect was observed.

To detect the non-reciprocal responses, we applied an AC source-drain current $I_{\textrm{AC}}$ (frequency $f=27$ Hz and amplitude 5 nA) mixed with a DC current $I_{\textrm{DC}}$ ($|I_{\textrm{DC}}|<20$ nA), letting $I=I_{\textrm{AC}}+I_{\textrm{DC}}$.
We then measured the $I_{\textrm{DC}}$ dependence of the differential resistance, which is $\frac{dV_{xx}}{dI}=R_{xx}+R_{xx}^{(2)}I$ and $\frac{dV_{xy}}{dI}=R_{xy}+R_{xy}^{(2)}I$. 
In this method, the differential resistance at $I_{\textrm{DC}}=0$ corresponds to the linear resistance ($R_{xx}$ and $R_{xy}$), and the slope of the $I_{\textrm{DC}}$ dependence corresponds to that of the non-reciprocal component ($R_{xx}^{(2)}$ and $R_{xy}^{(2)}$). Note that we could also detect the non-reciprocal responses by measuring the second harmonic ($27\times2=54$ Hz) voltage response. The measured first (and second) harmonic voltage of the longitudinal and transverse resistance correspond to $R_{xx}$ ($R_{xx}^{(2)}$) and $R_{xy}$ ($R_{xy}^{(2)}$), respectively. We have also used this method to evaluate the non-reciprocal signal and obtained the same results.

Figure \ref{sample}(c) shows the $I_{\textrm{DC}}$ dependence of the differential resistance $\frac{dV_{xx}}{dI}$ and $\frac{dV_{xy}}{dI}$ within $\pm 20$ nA. At the quantum Hall plateau ($\nu=1$), no $I_{\textrm DC}$ dependence was observed, and the resistance remained constant, meaning that the transport is perfectly linear within the measured range of the current.
Now, it is a surprise that $\frac{dV_{xx}}{dI}$ and $\frac{dV_{xy}}{dI}$ exhibited a linear dependence on $I_{\textrm{DC}}$ in the out-of-plateau regime ($\nu=0.7$) as shown in Fig. \ref{sample}(d). This means that a finite non-reciprocity emerges in this regime. 
The relative amplitude of the nonreciprocal signal compared to the linear component is $\frac{R_{xx}^{(2)}I^2}{R_{xx}I}\leq\frac{2.5 \Omega/\textrm{nA}\times (5 \textrm{nA})^{2}}{4 k\Omega\times 5 \textrm{nA}}\simeq0.3\%$ and $\frac{R_{xy}^{(2)}I^2}{R_{xy}I}\leq\frac{2.5 \Omega/\textrm{nA}\times (5 \textrm{nA})^{2}}{35 k\Omega\times 5 \textrm{nA}}\simeq0.03\%$, meaning that the linear component is still dominant.
From the fact that no $I_{\textrm{DC}}$ dependence was observed at the plateau, we can say that the observed nonlinearity is irrelevant to contact resistance nor the breakdown of the quantum Hall effect.
We have also done a control experiment using a sample without back gate, in which no $I_{\textrm DC}$ dependence was observed (see Appendix \ref{section:control}). This suggests that the existence of the back gate is essential to observe the non-reciprocity.

We characterize the behavior of the non-reciprocal response ($R_{xx}^{(2)}$ and $R_{xy}^{(2)}$) with respect to the magnetic field and the contact configurations. In our sample, there are two configurations for measuring $V_{xx}$. One is along the top channel of the sample (from contact 1 to 2 in Fig. \ref{sample}(a)), and the other is along the bottom (from contact 3 to 4 in Fig. \ref{sample}(a)).
As long as we deal with the linear response, there is no configuration dependence in the results. However, the configuration actually matters in the non-reciprocal response, as we see below.
In Fig. \ref{asymmetry}(a), we show two traces of $R_{xx}^{(2)}$ for the two contact configurations, namely top and bottom (see the inset in the Fig \ref{asymmetry}(a)). 
Consistent with the results in Fig. \ref{sample}(c), $R_{xx}^{(2)}$ takes zero at the quantum Hall plateau regime and a finite value outside the plateau. Unlike $R_{xx}$, $R_{xx}^{(2)}$ is not symmetric with the reversal of the magnetic field ($B\rightarrow-B$). Moreover, it clearly depends on the configurations and there seems to be a certain symmetric correlation between the two datasets. 

There are also two configurations for measuring $V_{xy}$. One is on the left side of the sample (from contact 1 to 3 in Fig. \ref{sample}(a)), and the other is on the right side (from contact 2 to 4 in Fig. \ref{sample}(a)).
Figure \ref{asymmetry}(b) shows the magnetic field and configuration dependence of $R_{xy}^{(2)}$. Similar to the previous case, a peculiar symmetry in the magnetic field and the configuration dependence was also observed.
The actual formulations of the magnetic field and configuration symmetry will be discussed and derived in the following section.
We also measured $R_{xx}^{(2)}$ and $R_{xy}^{(2)}$ with different back gate voltages of $V_{\textrm{G}}=-$1 V ($n=1.93\times10^{11}$ /cm$^{2}$) and $0$ V ($n=2.50\times10^{11}$ /cm$^{2}$). The obtained magnetic field dependence is essentially the same as the ones shown in Fig.\ref{asymmetry} (see Appendix \ref{section:Vbg dep} for data).
The observation of finite non-reciprocal responses and their magnetic field and configuration symmetries are the central experimental findings in this Article.

\begin{figure}[ht]
\centering
\includegraphics[scale=0.4]{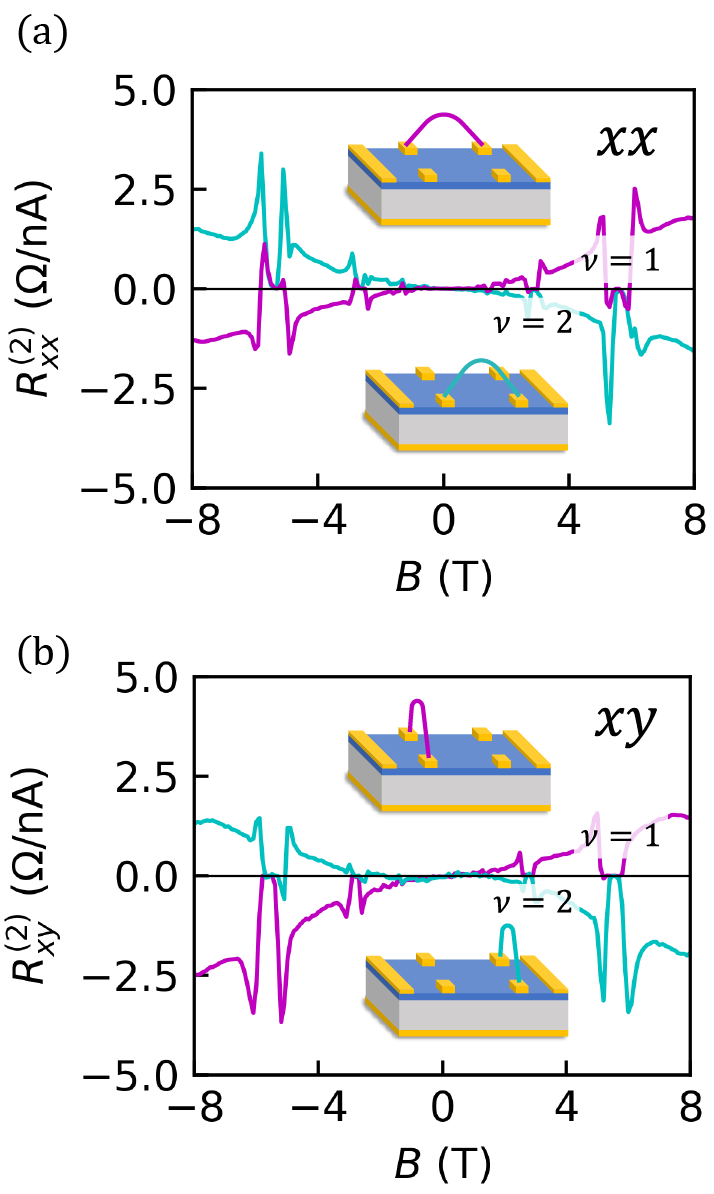}
\caption{Magnetic field dependence of the non-reciprocal component (a) $R_{xx}^{(2)}$ (magenta: top, cyan: bottom) and (b) $R_{xy}^{(2)}$ (magenta: left, cyan: right). Insets show contact configurations.}
\label{asymmetry}
\end{figure}
\section{Analysis}
Now, we discuss the origin of the non-reciprocal responses as well as the field and the configuration symmetries. The observation of the non-reciprocal response infers that the inversion symmetry is broken in the sample. We suggest that it is caused by the capacitive coupling between the 2DEG and the back gate. Our model assumes that the carrier density is influenced by both the back gate voltage and the Hall voltage, which alters the potential difference between the 2DEG and back gate. This creates a gradient of carrier density across and along the sample \cite{PhysRevB.73.235333,PhysRevLett.73.3278}, breaking the inversion symmetry and leading to non-reciprocal behavior \cite{PhysRevLett.105.146802}.

\subsection{Model}
\begin{figure*}[ht]
\centering
\includegraphics[scale=0.32]{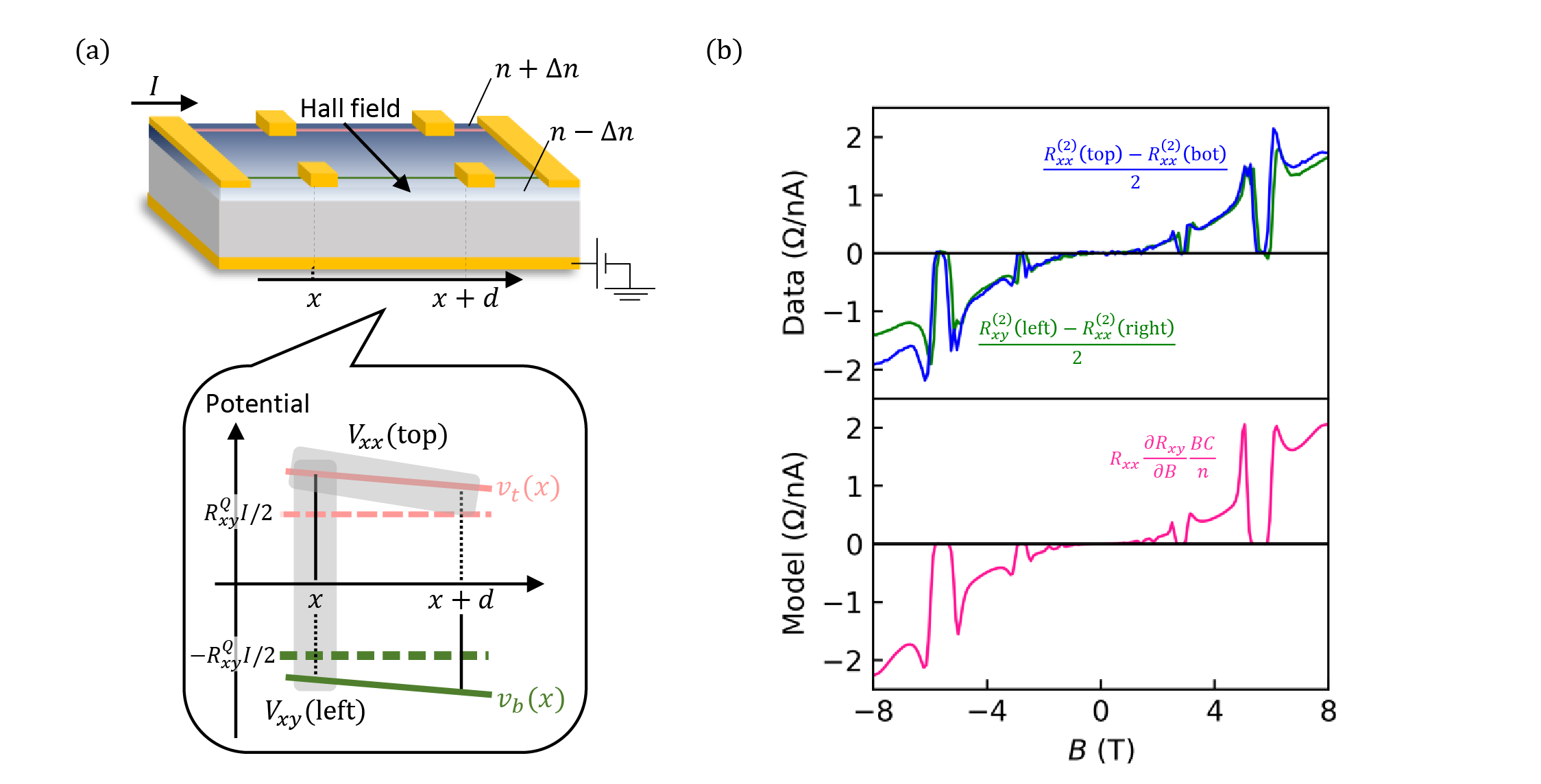}
\caption{(a) Schematic summary of the model. Due to the Hall effect, a potential gradient in the transverse direction of the sample is created. This modulates the potential between the 2DEG and the back gate space-dependent, modulating the carrier density in space ($n\pm\Delta n$ at the sample edges are shown). (Popup) Schematic of the Hall voltage along the $x$ (longitudinal) axis of the sample. Solid (dotted) lines show the potential profile outside (inside) the plateau. Red and blue lines correspond to the top and bottom channels, respectively. Potential difference that corresponds to $V_{xx}$ and $V_{xy}$ are indicated by red and blue rectangles. (b) Comparison of the data from the experiment $\frac{R_{xx}^{(2)}(\textrm{top})-R_{xx}^{(2)}(\textrm{bot})}{2}$ (blue line) and $\frac{R_{xy}^{(2)}(\textrm{left})-R_{xy}^{(2)}(\textrm{right})}{2}$ (green line) and the model $R_{xx}\frac{\partial R_{xy}}{\partial B}\frac{BC}{n}$ (pink line) setting $C=$10 pF/cm$^2$.}
\label{model}
\end{figure*}

In the quantum Hall plateau regime, the current is carried by the edge channels and no dissipation nor potential drop occurs. Therefore, the potential distribution would look like the dotted horizontal lines shown in Fig. \ref{model}(a). $v_{t}(x, B)$ and $v_{b}(x, B)$ are the potential along the top and the bottom channel. In the Figure, $R_{xy}^{Q}=\frac{h}{\nu e^2}$ is the Hall resistance at the quantum Hall plateau. The contacts along the top channel separated by a distance $d$ see the voltage $V_{xx}(\textrm{top})$, which is the potential difference between $v_{t}(x)$ and $v_{t}(x+d)$.
Then, $V_{xx}(\textrm{top})$ is given by $V_{xx}(\textrm{top})=v_{t}(x)-v_{t}(x+d)=R_{xy}^{Q}I/2-R_{xy}^{Q}I/2=0$. Similarly, the Hall voltage $V_{xy}(\textrm{left})$ is given by the voltage difference between the two contacts on the left or right side of the sample, $V_{xy}(\textrm{left})=v_{t}(x)-v_{b}(x)=R_{xy}^{Q}I/2-(-R_{xy}^{Q}I/2)=R_{xy}^{Q}I$. Note that both $V_{xx}(\textrm{top})$ and $V_{xy}(\textrm{left})$ are linear in $I$. This reflects the experimental observation that there is no non-linearity in the plateau regime.

Next, we consider the situation outside the plateau. The current is carried not only by the edge channel but also by the bulk, giving rise to a finite dissipation and a potential drop.
The position-dependent potential can be written as $v_{t}(x)=\rho_{xy}(\textrm{top},x,n(x))I$, where resistivity $\rho_{xy}(\textrm{top},x,n(x))$ is that of top channel as a function of $x$ and $n(x)$. Here, the key assumption is that the carrier density is also a function of $x$, making $n(x)$. Considering that the potential difference between the 2DEG and the back gate determines the carrier density, $n(x)$ can be written such as $n(x)=N+C\left[V_{\textrm{G}}+v_{t}(x)\right]=n_{0}+Cv_{t}(x)$. Here, $N$ is the carrier density without back gate voltage nor the injection current, $C$ is the capacitance between the 2DEG and the back gate, and $n_{0}=N+CV_{\textrm{G}}$.

Hereafter, we briefly sketch the outline of the derivation of the non reciprocal responses. See Appendix \ref{section:derivation} for detailed derivation.
We consider the longitudinal voltage along the top channel $V_\textrm{xx}(\textrm{top})$ as an example.
The linear term of $V_\textrm{xx}(\textrm{top})$ comes from the spatial dependence of the resistivity. Assuming a constant carrier density of $n$, $V_{xx}(\textrm{top})=\rho_{xy}\left(\textrm{top},x,n\right)I-\rho_{xy}\left(\textrm{top},x+d,n\right)I=-\frac{\partial \rho_{xy}(\textrm{top},x,n)}{\partial x}dI\propto I$. As we see in Appendix \ref{section:derivation}, the coefficient of $I$ corresponds to $R_{xx}$.
The nonreciprocal term of $V_{xx}(\textrm{top})$ can be obtained considering the spatial dependence of carrier density $n(x)$. 
$V_\textrm{xx}(\textrm{top})=\rho_{xy}(\textrm{top},n(x))I-\rho_{xy}\left(\textrm{top},n(x+d\right))I\simeq\frac{\partial\rho_{xy}}{\partial n}[n(x)-n(x+d)]I=\frac{\partial\rho_{xy}}{\partial n}C[v_{t}(x)-v_{t}(x+d)]I$.
Here, we assume that the amount of modulation of the carrier density due to the injection current is small enough compared to the original carrier density ($Cv_{t}\left(x\right)\ll n_{0}$).
Using $v_{t}(x)-v_{t}(x+d)=R_{xx}(\textrm{top})I$, we obtain $V_{xx}(\textrm{top})=R_{xx}(\textrm{top})\frac{\partial\rho_{xy}}{\partial n}CI^2\propto I^2$.
Similarly, expressions of $R_{xx}^{(2)}$ and $R_{xy}^{(2)}$ for all configurations are obtained (see Appendix \ref{section:derivation} for detailed derivation).

\subsection{Magnetic field and configuration symmetry}
Using the model shown above, we derive the magnetic field and configuration symmetries of the non-reciprocal resistances. 
For $R_{xx}^{(2)}$ and $R_{xy}^{(2)}$, one obtains the relation below.
\begin{equation}
\begin{split}
R_{xx}^{(2)}(\textrm{top},B)&=R_{xx}^{(2)}(\textrm{bot},-B).\\
R_{xy}^{(2)}(\textrm{left},B)&=-R_{xy}^{(2)}(\textrm{left},-B)\\
R_{xy}^{(2)}(\textrm{right},B)&=-R_{xy}^{(2)}(\textrm{right},-B).
\end{split}  
\label{TB_App}
\end{equation}
These outcomes on the commutation in magnetic field and configuration in $R_{xx}^{(2)}$ and the anti-symmetry in $R_{xy}^{(2)}$ are in approximate agreement with the experimental observation in Fig \ref{asymmetry}.

According to the model, $R_{xy}^{(2)}$ and $R_{xx}^{(2)}$ should also be correlated to each other.

\begin{equation}
R_{xx}^{(2)}(\textrm{top})-R_{xx}^{(2)}(\textrm{bot})=R_{xy}^{(2)}(\textrm{left})-R_{xy}^{(2)}(\textrm{right})
\label{LRTB0}
\end{equation}
\begin{equation}
\begin{split}
R_{xx}^{(2)}(\textrm{top})-R_{xx}^{(2)}(\textrm{bot})&=R_{xy}^{(2)}(\textrm{left})-R_{xy}^{(2)}(\textrm{right})\\
&=2CR_{xx}\frac{\partial R_{xy}}{\partial n_{0}}\\
&=-\frac{2CB}{n}R_{xx}\frac{\partial R_{xy}}{\partial B}.
\end{split}  
\label{LRTB}
\end{equation}

In the last transformation, Euler's chain rules $ \left(\frac{\partial R_{xy}}{\partial n_{0}}\right)_{B}=-\left(\frac{\partial B}{\partial n_{0}}\right)_{R_{xy}}\left(\frac{\partial R_{xy}}{\partial B}\right)_{n}=-\frac{B}{n}\left(\frac{\partial R_{xy}}{\partial B}\right)_n$ are used  \cite{PhysRevLett.73.3278,Pan2005}.
Equation (\ref{LRTB0}) illustrates that the potential drop across the top and right channels are equivalent to that across the bottom and left channels. This equivalence should be consistently maintained, independent of the specificities of the model.

Now, we compare the calculation with the experimental data shown in Fig. \ref{model} (c). The two traces in the top panel demonstrates a correspondence between $R_{xy}^{(2)}(\textrm{left})-R_{xy}^{(2)}(\textrm{right})$ and $R_{xx}^{(2)}(\textrm{top})-R_{xx}^{(2)}(\textrm{bot})$ as predicted in Eq. (\ref{LRTB0}).
Note that this relation is not exactly satisfied, where the blue and green curves are not exactly on each other. This is because the four components were not measured simultaneously. Initially, $R_{xy}^{(2)}(\textrm{left})$ and $R_{xx}^{(2)}(\textrm{top})$ were measured and after changing the terminal configurations, $R_{xy}^{(2)}(\textrm{right})$ and $R_{xx}^{(2)}(\textrm{bottom})$ were measured subsequently. 
Therefore, the two measurements are not perfectly identical due to unavoidable factors such as fluctuations in the magnetic field, subtle temperature variations, and minor changes in the gate voltage.

Moreover, the expectation from the model (Eq. (\ref{LRTB})) and the data also agree with each other, as shown in the bottom panel of Fig. \ref{model}(c). In the bottom panel of Fig. \ref{model}(c), the capacitance $C$ is assumed to be 300 nF/cm$^2$. These results show that our model captures the behavior of the non-reciprocal responses very well. However, the capacitance we assumed to obtain a quantitative agreement ($\sim$300 nF/cm$^{2}$) is by a factor of $\sim$30 bigger than the value estimated with other methods (9.6 nF/cm$^2$ by carrier density and 12 nF/cm$^2$ by direct measurement). 
The discrepancy between the data and the model may be due to the effect of quantum capacitance. The capacitance used in the plot only considered the geometric capacitance at zero magnetic field, while the total capacitance ($C_\textrm{tot}$) is composed of both geometric capacitance ($C_\textrm{geo}$) and quantum capacitance ($C_Q$) such that $\frac{1}{C_\textrm{tot}}=\frac{1}{C_\textrm{geo}}+\frac{1}{C_{Q}}$. The quantum capacitance is given by $C_Q = \frac{dn}{d\mu}$, where $n$ is the carrier density and $\mu$ is the chemical potential. In the vicinity of the quantum Hall regime, where the non-reciprocal responses were observed, $C_Q$ can be negative, resulting in $C_\textrm{tot}$ being greater than $C_\textrm{geo}$ \cite{PhysRevB.50.1760,PhysRevLett.68.674}. Accounting for this effect could explain the discrepancy between the data and the model, but further studies are needed to determine the microscopic origin of the non-reciprocity.
\section{Conclusion}
In summary, we investigated the non-reciprocal transport response in the quantum Hall regime next to the plateau regions
in high magnetic field. We also found that these responses obey certain symmetry relations and are reconstructed from linear response coefficients. Our phenomenological model based on the capacitive coupling between the 2DEG and the back gate also supports the observation, whereas the origin of a quantitative discrepancy remains unsolved. These results suggest a universal mechanism of obtaining non-reciprocal responses in gated devices.

\begin{acknowledgments}
We are grateful for the technical support from Peter Maerki, Thomas Baehler, and the ETH FIRST cleanroom facility staff. 

The authors acknowledge financial support from Eidgenössische Technische Hochschule Zürich (ETH Z\"urich) and the Swiss National Science Foundation via the National Center of Competence in Research Quantum Science and Technology (NCCR QSIT).
\end{acknowledgments}
\setcounter{equation}{0}
\setcounter{figure}{0}
\renewcommand{\theequation}{A.\arabic{equation}}
\renewcommand{\thefigure}{A.\arabic{figure}}

\revappendix
\section{Control experiment}
\label{section:control}
\begin{figure}[h]
\centering
\includegraphics[scale=0.25]{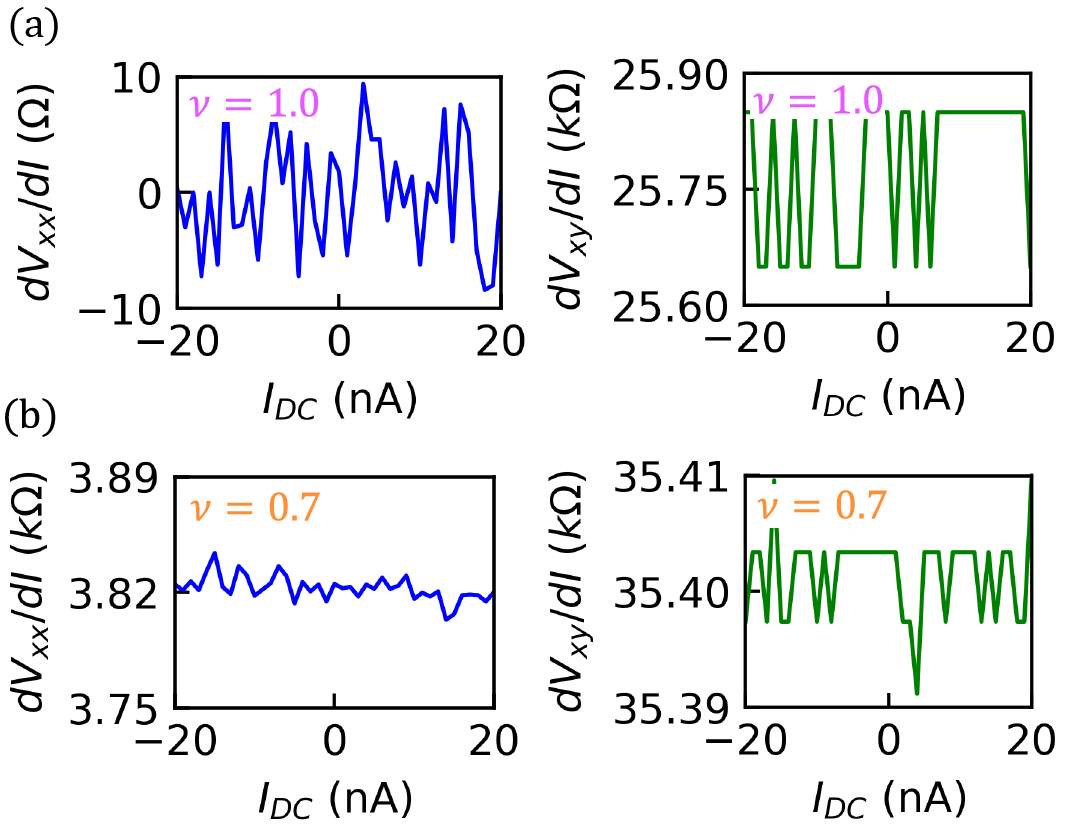}
\caption{DC current $I_{\textrm DC}$ dependence of the differential resistances in plateau ($\nu=1$) regime ((a)left panel: $\frac{dV_{xx}}{dI}$, right panel: $\frac{dV_{xy}}{dI}$) and outside plateau ($\nu=0.7$) regime ((b)left panel: $\frac{dV_{xx}}{dI}$, right panel: $\frac{dV_{xy}}{dI}$).}
\label{control}
\end{figure}

We have conducted a control experiment using a GaAs/GaAlAs without back gate ($n=$ 1.2$\times10^{11}$/cm$^{2}$). We measured $\frac{dV_{xx}}{dI}$ and $\frac{dV_{xy}}{dI}$ as a function of $I_{\textrm{DC}}$ at $\nu=$0.7 and 1 as shown in Fig. \ref{control}.
Here, no $I_{\textrm{DC}}$ dependence was observed. This result suggests that the presence of back gate is essential to generate non-reciprocal responses.

\section{Back gate voltage dependence}
\label{section:Vbg dep}
\begin{figure}[h]
\centering
\includegraphics[scale=0.25]{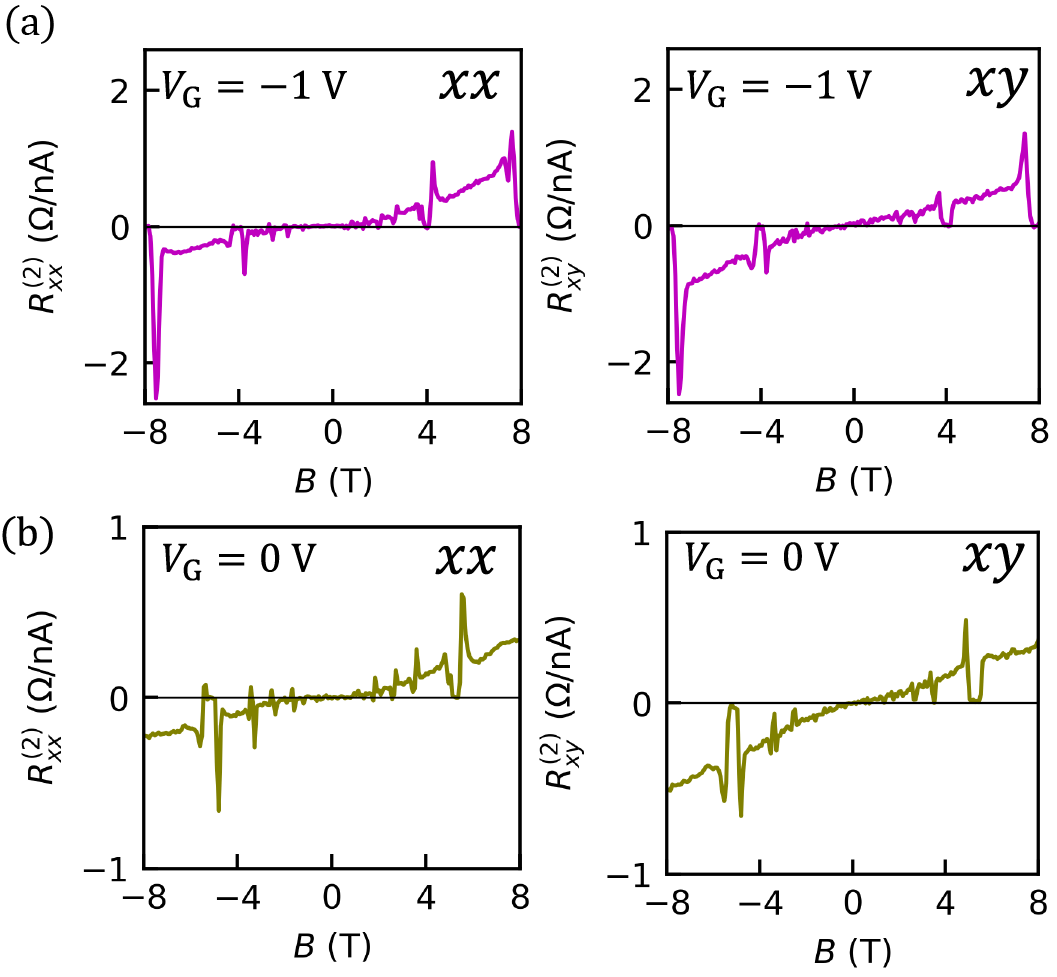}
\caption{Magnetic field dependence of the non-reciprocal component. (a) $V_{\textrm{G}}=-1$ V (left: $R_{xx}^{(2)}$(top), right: $R_{xy}^{(2)}$(left))(b) $V_{\textrm{G}}=0$ V (left: $R_{xx}^{(2)}$(top), right: $R_{xy}^{(2)}$(left)).}
\label{backgateDep}
\end{figure}

Figure \ref{backgateDep} shows the magnetic field dependence of $R_{xx}^{(2)}$ and $R_{xy}^{(2)}$ at two different back gate voltages of $-1$ V ($n=1.93\times10^{11}$ /cm$^{2}$) and $0$ V ($n=2.50\times10^{11}$ /cm$^{2}$). 
The same behavior as seen in Fig. \ref{asymmetry} (zeros at the quantum Hall plateau and antisymmetries in magnetic field) is observed.

\section{Capacitance measurement}
\label{section:capacitance}
We have estimated the capacitance between 2DEG and back gate by two means. First, we measured a classical Hall effect at low magnetic field ($<$0.5 T) with which we deduce the carrier densities as a function of back gate. By fitting $n=N+CV_{\textrm{G}}$ ($N$ is the density at zero gate voltage), we estimated the capacitance $C\sim9.6$ nF/cm$^2$.
Second, we connected a current-to-voltage ($IV$) converter to 2DEG and grounded back gate. By applying an AC voltage to 2DEG, we measured the out-of-phase component of the current flowing through the $IV$ converter. This circuit can be considered as measuring a charging current of a capacitor (with negligible resistance and inductance) under an AC voltage application, from which we deduce the capacitance $C\sim12$ nF/cm$^2$.
\begin{figure}[h]
\includegraphics[scale=0.35]{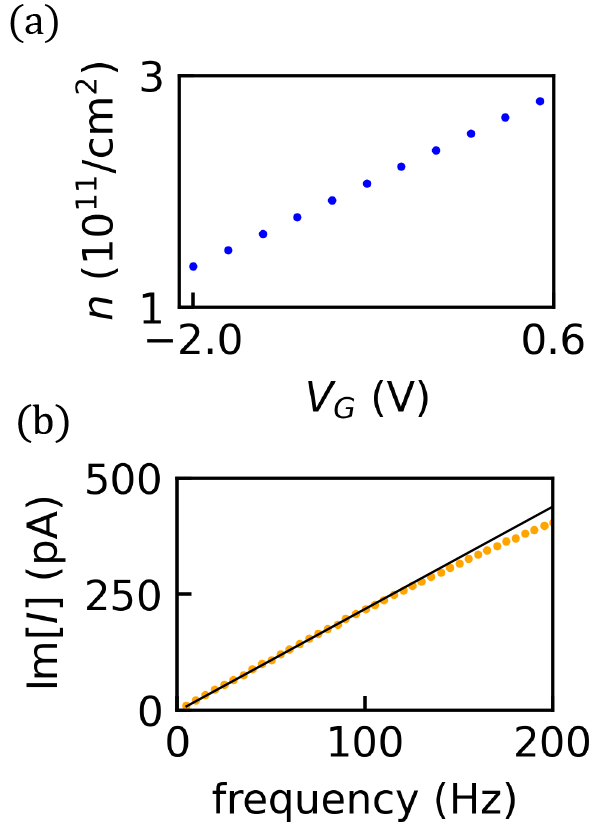}
\caption{(a) Back gate dependence of carrier density. (b) Frequency dependence of the out-of-phase component of the charging current across 2DEG and back gate.}
\label{Capacitances}
\end{figure}

\section{Derivation of the model}
\label{section:derivation}
The configuration, position, and carrier density-dependent potential is expressed as $v_{t}(x)=\rho_{xy}(\textrm{top},x,n(x))I$. As discussed in the main script, the key assumption is that the carrier density is also a function of $x$, making $n(x)$. Considering that the carrier density is determined by the potential difference between the 2DEG and the back gate, $n(x)$ can be written as $n(x)=N+C\left[V_{\textrm{BG}}+v_{t}(x)\right]=n_{0}+Cv_{t}(x)$. Here, $N$ is the carrier density without back gate voltage and the injection current, $n_{0}=N+CV_{\textrm{BG}}$ and $C$ is the capacitance between the 2DEG and the back gate. Then, $v_{t}(x)$ can be expressed as below.
\begin{equation}
\begin{split}
v_{t}(x)&=\rho_{xy}\left(\textrm{top},x,n\left(x\right)\right)I\\
&\simeq\rho_{xy}\left(\textrm{top},x,n_{0}\right)I+\frac{\partial\rho_{xy}}{\partial n_{0}}Cv_{t}\left(x\right)I.
\end{split}
\label{Tx}
\end{equation}
Here, we assumed that the amount of modulation of the carrier density due to the injection current is small enough compared to the original carrier density ($Cv_{t}\left(x\right)\ll n_{0}$).
The above equation can be solved with $v_{t}(x)$.
\begin{equation}
\begin{split}
v_{t}(x)&\simeq\frac{\rho_{xy}\left(\textrm{top},x,n_{0}\right)I}{1-\frac{\partial\rho_{xy}}{\partial n_{0}}CI}\\
&\simeq\rho_{xy}^{\textrm{t}}(x)I+\rho_{xy}^{\textrm{t}}(x)\frac{\partial\rho_{xy}^{\textrm{t}}(x)}{\partial n_{0}}CI^{2},
\end{split}
\label{vt(x)}
\end{equation}
where $\rho_{xy}^{\textrm{t}}(x)=\rho_{xy}\left(\textrm{top},x,n_{0}\right)$. We assumed that the amount of potential modulation due to the injection current is small enough compared to the total potential ($\frac{\partial\rho_{xy}^{\textrm{t}}(x)}{\partial n_{0}}Cv_{t}(x)I\ll v_{t}(x)I$), which also means that the first term of Eq. (\ref{Tx}) is much larger than the second term.
We assume that the potential changes linearly in space $v_{t}(x+d)= v_{t}(x)+\frac{\partial v_{t}(x)}{\partial x}d+o(d^{2})$ and obtain the expression of $v_{t}(x+d)$.
\begin{equation}
\begin{split}
v_{t}(x+d)&\simeq v_{t}(x)+\frac{\partial\rho_{xy}^{t}}{\partial x}dI\\
&+\left(\frac{\partial\rho_{xy}^{t}}{\partial x}\frac{\partial\rho_{xy}^{t}}{\partial n_{0}}+\rho_{xy}^{t}\frac{\partial^{2}\rho_{xy}^{\textrm{t}}}{\partial x \partial n_{0}}\right)\times dCI^2
\end{split}
\label{vt(x+d)}   
\end{equation}

One can obtain the expression for $v_{b}(x)$ and $v_{b}(x+d)$ by exchanging $\rho_{xy}^{\textrm{t}}(x)$ with $\rho_{xy}^{\textrm{b}}(x)$ in Eq. (\ref{vt(x)}) and (\ref{vt(x+d)}).

The expressions for the linear and the nonlinear resistances are obtained by calculating $V_{xx}(\textrm{top})=v_{t}(x)-v_{t}(x+d)$, $V_{xx}(\textrm{bot})=v_{b}(x)-v_{b}(x+d)$, $V_{xy}(\textrm{left})=v_{t}(x)-v_{b}(x)$, and $V_{xy}(\textrm{right})=v_{t}(x+d)-v_{b}(x+d)$.
In each expression, the term proportional to $I$ and $I^{2}$ are assigned to the linear ($R_{xx}$ and $R_{xy}$) and the non-reciprocal ($R_{xx}^{(2)}$ and $R_{xy}^{(2)}$) resistances that are observed experimentally.
\begin{equation}
R_{xx}=-\frac{\partial\rho_{xy}^{\textrm{t}}(x)}{\partial x}d.
\label{Rxx}
\end{equation}
\begin{equation}
R_{xy}=\rho_{xy}^{\textrm{t}}(x)-\rho_{xy}^{\textrm{b}}(x).
\label{Rxy}
\end{equation}
\begin{equation}
\begin{split}
R_{xx}^{(2)}(\textrm{top})&=-2\Biggl[\frac{\partial\rho_{xy}^{\textrm{t}}(x)}{\partial x}\frac{\partial\rho_{xy}^{\textrm{t}}(x)}{\partial n_{0}}+\rho_{xy}^{\textrm{t}}(x)\frac{\partial^{2}\rho_{xy}^{\textrm{t}}(x)}{\partial x\partial n_{0}}\Biggr]\\
&\times dC.
\end{split}
\label{Rxx(top)}
\end{equation}
\begin{equation}
\begin{split}
R_{xy}^{(2)}(\textrm{left})&=2\left(\rho_{xy}^{\textrm{t}}(x)\frac{\partial\rho_{xy}^{\textrm{t}}(x)}{\partial n_{0}}-\rho_{xy}^{\textrm{b}}(x)\frac{\partial\rho_{xy}^{\textrm{b}}(x)}{\partial n_{0}}\right)C.
\end{split}
\label{Rxy(left)}
\end{equation}
\begin{equation}
\begin{split}
R_{xx}^{(2)}(\textrm{bot})&=-2\Biggl[\frac{\partial\rho_{xy}^{\textrm{b}}(x)}{\partial x}\frac{\partial\rho_{xy}^{\textrm{b}}(x)}{\partial n_{0}}+\rho_{xy}^{\textrm{b}}(x)\frac{\partial^{2}\rho_{xy}^{\textrm{b}}(x)}{\partial x\partial n_{0}}\Biggr]\\
&\times dC.
\end{split}
\label{Rxx(bot)}
\end{equation}
\begin{equation}
\begin{split}
R_{xy}^{(2)}(\textrm{right})&=\biggl(\frac{\partial\rho_{xy}^{\textrm{t}}(x)}{\partial x}\frac{\partial\rho_{xy}^{\textrm{t}}(x)}{\partial n_{0}}+\rho_{xy}^{\textrm{t}}(x)\frac{\partial^{2}\rho_{xy}^{\textrm{t}}(x)}{\partial x\partial n_{0}}\biggr)\\
&\times 2dC\\
&-\biggl(\frac{\partial\rho_{xy}^{\textrm{b}}(x)}{\partial x}\frac{\partial\rho_{xy}^{\textrm{b}}(x)}{\partial n_{0}}+\rho_{xy}^{\textrm{b}}(x)\frac{\partial^{2}\rho_{xy}^{\textrm{b}}(x)}{\partial x\partial n_{0}}\biggr)\\
&\times 2dC\\
&+2C\left(\rho_{xy}^{\textrm{t}}(x)\frac{\partial\rho_{xy}^{\textrm{t}}(x)}{\partial n_{0}}-\rho_{xy}^{\textrm{b}}(x)\frac{\partial\rho_{xy}^{\textrm{b}}(x)}{\partial n_{0}}\right).
\end{split}
\label{Rxy(right)}
\end{equation}
Using the model shown above, we derive the magnetic field and configuration symmetries of the non-reciprocal resistances. 
To derive the symmetry relations, we first get three useful relations to discuss the magnetic field symmetry. 
First, the reversal of the magnetic field is equivalent to exchanging the top and the bottom channels.
\begin{equation}
\rho_{xy}^{\textrm{t}}(B)=\rho_{xy}^{\textrm{b}}(-B)
\label{TBsymmetry}
\end{equation}
Second, using the fact that $R_{xx}(\textrm{top})=R_{xx}(\textrm{bot})$, 
\begin{equation}
\begin{split}
\frac{\partial\rho_{xy}^{\textrm{t}}}{\partial x}=\frac{\partial\rho_{xy}^{\textrm{b}}}{\partial x}.
\end{split}
\label{dx symmetry}
\end{equation}
Third, by differentiating Eq. (\ref{Rxy}) with $n$, one obtains
\begin{equation}
\begin{split}
\frac{\partial R_{xy}}{\partial n_{0}}=\frac{\partial\rho_{xy}^{\textrm{t}}}{\partial n_{0}}-\frac{\partial\rho_{xy}^{\textrm{b}}}{\partial n_{0}}.
\end{split}
\label{dn symmetry}
\end{equation}

By using (\ref{Rxy(left)}), (\ref{Rxy(right)}), and (\ref{TBsymmetry}), one obtains the relations below.
\begin{equation}
\begin{split}
R_{xy}^{(2)}(\textrm{left},B)&=-R_{xy}^{(2)}(\textrm{left},-B)\\
R_{xy}^{(2)}(\textrm{right},B)&=-R_{xy}^{(2)}(\textrm{right},-B).
\end{split}
\label{symmetryRxy_App}
\end{equation}
This means that the $R_{xy}^{(2)}(\textrm{left})$ is antisymmetric in $B$. The same goes for $R_{xy}^{(2)}(\textrm{right})$. 
Also, Eq. (\ref{Rxx}) and (\ref{Rxx(top)}) leads to another symmetry relation for $R_{xx}^{(2)}$.
\begin{equation}
  R_{xx}^{(2)}(\textrm{top},B)=R_{xx}^{(2)}(\textrm{bot},-B).
\label{TB}
\end{equation}
The relation between $R_{xy}^{(2)}$ and $R_{xx}^{(2)}$ are also obtained by combining Eq. (\ref{Rxx(top)})-(\ref{Rxy(right)}).
\begin{equation}
R_{xy}^{(2)}(\textrm{left})-R_{xy}^{(2)}(\textrm{right})=R_{xx}^{(2)}(\textrm{top})-R_{xx}^{(2)}(\textrm{bot}).  
\end{equation}

We have seen the symmetry relations within $R_{xy}^{(2)}$ and $R_{xx}^{(2)}$, and we can further derive the connection between them using Eq. (\ref{Rxx(top)})-(\ref{Rxy(right)}), and (\ref{TB}).
\begin{equation}
\begin{split}
R_{xy}^{(2)}(\textrm{left})-R_{xy}^{(2)}(\textrm{right})&=R_{xx}^{(2)}(\textrm{top})-R_{xx}^{(2)}(\textrm{bot})\\
&=2CR_{xx}\frac{\partial R_{xy}}{\partial n_{0}}\\
&=-\frac{2CB}{n}R_{xx}\frac{\partial R_{xy}}{\partial B}.
\end{split}  
\label{LRTB_App}
\end{equation}
The transformation from the first line to the second line goes like below. By plugging Eq. (\ref{dx symmetry}) and (\ref{dn symmetry}) into Eq. (\ref{Rxx(top)})
\begin{equation}
    \begin{split}
        R_{xx}^{(2)}(\textrm{top})&=-2dC\left(\frac{\partial\rho_{xy}^{\textrm{b}}(x)}{\partial x}\frac{\partial\rho_{xy}^{\textrm{t}}(x)}{\partial n_{0}}+\rho_{xy}^{\textrm{b}}(x)\frac{\partial}{\partial x}\frac{\partial \rho_{xy}^{\textrm{b}}}{\partial n_{0}}\right)\\
        &+2dC\left(\rho_{xy}^{\textrm{b}}(x)\frac{\partial}{\partial x}\frac{\partial \rho_{xy}^{\textrm{b}}}{\partial n_{0}}\right)\\
        &+2dC\left[\rho_{xy}^{\textrm{t}}(x)\frac{\partial }{\partial x}\left(\frac{\partial R_{xy}}{\partial n_{0}}+\frac{\partial \rho_{xy}^{b}}{\partial n_{0}}\right)\right]\\
        &=R_{xx}^{(2)}(\textrm{bot})-2dC\left[\left(\rho_{xy}^{\textrm{t}}-\rho_{xy}^{\textrm{b}}\right)\frac{\partial }{\partial x}\left(\frac{\partial \rho_{xy}^{\textrm{b}}}{\partial n_{0}}\right)\right]\\
        &-2dC\left(\frac{\partial \rho_{xy}^{\textrm{b}}}{\partial x}\frac{\partial R_{xy}}{\partial n_{0}}\right)\\
        &=R_{xx}^{(2)}(\textrm{bot})+2CR_{xx}\frac{\partial R_{xy}}{\partial n_{0}}
    \end{split}
\label{Rxx_transform1}
\end{equation}

In the transformation from the second line to the third line in Eq. (\ref{LRTB}), Euler's chain rules are applied for a given function of $R(n,B)$ \cite{PhysRevLett.73.3278,Pan2005} that is $ \left(\frac{\partial R_{xy}}{\partial n_{0}}\right)_{B}=-\left(\frac{\partial B}{\partial n_{0}}\right)_{R_{xy}}\left(\frac{\partial R_{xy}}{\partial B}\right)_{n}=-\frac{B}{n}\left(\frac{\partial R_{xy}}{\partial B}\right)_n$.

\bibliography{QHE_cite}

\end{document}